\documentclass[pra,aps,twocolumn,floatfix,10pt]{revtex4-1}

\usepackage{amsmath}
\usepackage{multirow}
\usepackage{graphicx}
\usepackage{subfigure}
\usepackage{epstopdf}
\usepackage{threeparttable}
\usepackage{tabularx}
\usepackage{hyperref}

\begin{document}

\title{Static field-gradient polarizabilities of small atoms and molecules in finite temperature}

\author{Juha Tiihonen}
\email{tiihonen@iki.fi}
\affiliation{Laboratory of Physics, Tampere University of Technology, P.O.~Box 692, FI-33101 Tampere, Finland}
\author{Ilkka Kyl\"anp\"a\"a}
\altaffiliation{Present address:\\Materials Science and Technology Division, Oak Ridge National Laboratory, Oak Ridge, Tennessee 37831, USA}
\affiliation{Laboratory of Physics, Tampere University of Technology, P.O.~Box 692, FI-33101 Tampere, Finland}
\author{Tapio T. Rantala}
\affiliation{Laboratory of Physics, Tampere University of Technology, P.O.~Box 692, FI-33101 Tampere, Finland}


\date{\today}

\begin{abstract}
In this work, we propose new field-free estimators for static field-gradient polarizabilities in finite temperature PIMC simulation. Namely, dipole--quadrupole polarizability $A$, dipole--dipole--quadrupole polarizability $B$ and quadrupole--quadrupole polarizability $C$ are computed for several up to two-electron systems: H, H$^-$, He, Li$^+$, Be$^{2+}$, Ps$_2$, PsH, H$_2^+$, H$_2$, H$_3^+$ and HeH$^+$. We provide complementary data for ground state electronic properties within the adiabatic approximation, and demonstrate good agreement with available values in the literature. More importantly, we present fully non-adiabatic results from 50 K to 1600 K, which allow us to analyze and discuss strong thermal coupling and rovibrational effects in total field-gradient polarizabilities. These phenomena are most relevant but clearly overlooked, \emph{e.g.}, in the construction of modern polarizable force field models. However, our main purpose is demonstrating the accuracy and simplicity of our approach in a problem that is generally challenging.
\end{abstract}


\maketitle

\section{Introduction}

Computation of electric field response at quantum mechanical level -- polarizability -- is a fundamental problem in electronic structure theory. Approaching it from the first-principles is challenging but well motivated: polarizabilities have implications in many physical properties and modeling aspects, such as optical response, and atomic and molecular interactions. Method development and understanding of polarizability has been vast over the past several decades, but the main focus has always been on the bare ground state properties \cite{Buckingham2007,Maroulis2006a,Mitroy2010}. While the finite temperature regime is formally well established \cite{Bishop1990}, explicit results beyond the Born--Oppenheimer approximation are scarce. By introducing efficient polarizability estimators for the finite temperature path-integral Monte Calo method (PIMC), we are aiming to change that.

In our recent article \cite{Tiihonen2016a}, we proposed a scheme for estimating static dipole polarizabilities in a field-free PIMC simulation. This was an imminent improvement to our earlier finite-field approach \cite{Tiihonen2015}. The resulting properties, including substantial rovibrational effects, were those corresponding to an isolated molecule in low density gas. However, the dipole-induced polarizabilities only describe the effects of a uniform electric field.

In this work, we complement our tools by introducing similar estimators for the field-gradient polarizabilities. According to the definitions of Buckingham \cite{Buckingham2007}, the foremost properties are dipole--quadrupole polarizability $A$, dipole--dipole--quadrupole polarizability $B$ and quadrupole--quadrupole polarizability $C$. As the names suggest, they have direct consequence in treating the long-range interactions between atoms or molecules. There is emerging interest in, \emph{e.g.}, polarizable force field models \cite{Leontyev2011,Baker2015} and van der Waals coefficient formulae \cite{Tao2016} employing polarizabilities of all orders. However, the employed properties are often only electronic averages or fully empirical fits, while rovibrational coupling is completely overlooked. Here, we show that finite temperature has an immense effect on total molecular field-gradient polarizabilities.

At first, we present the analytic forms of the field-free PIMC estimators. After this we demonstrate their capability in a series of simulations for different small atoms, ions and molecules. The results are compared against values available in the literature. However, to the best of our knowledge, many of them are presented here for the first time. This is most pronounced in the non-adiabatic simulations, which include all rovibrational and electronic effects in finite temperature.

\section{Theory}
A perturbation caused by a uniform external electric field $F_\alpha$ and the field-gradient $F_{\alpha\beta}=(\nabla F_\alpha)_\beta$ gives the Hamiltonian as
\begin{equation}
\hat{H}^{(1)} = \hat{H}^{(0)} - \hat{\mu}_\alpha F_\alpha - \tfrac{1}{3} \hat{\Theta}_{\alpha \beta} F_{\alpha \beta} - \ldots,
\label{eq-hamiltonian}
\end{equation}
where $\hat{H}^{(0)}$ is the unperturbed Hamiltonian and $\hat{\mu}_\alpha$ and $\hat{\Theta}_{\alpha \beta}$ are the dipole and (traceless) quadrupole moment operators, respectively. Indices $\alpha, \beta, \gamma, \delta, \ldots$ refer to the Einstein summation of the combinations of $x$, $y$ and $z$. According to the Buckingham convention \cite{Buckingham2007}, the change in total energy is written as a perturbation expansion of coefficients
\begin{equation}
\begin{split}
E^{(1)} = & E^{(0)} - \mu_\alpha F_\alpha - \tfrac{1}{2}\alpha_{\alpha \beta} F_\alpha F_\beta - \tfrac{1}{6} \beta_{\alpha \beta \gamma} F_\alpha F_\beta F_\gamma \\ & - \tfrac{1}{24} \gamma_{\alpha \beta \gamma \delta} F_\alpha F_\beta F_\gamma F_\delta -\tfrac{1}{3} \Theta_{\alpha\beta} F_{\alpha\beta} \\ & - \tfrac{1}{3} A_{\gamma,\alpha\beta}F_\gamma F_{\alpha \beta} - \tfrac{1}{6} B_{\alpha\beta, \gamma \delta} F_\alpha F_\beta F_{\gamma \delta}  \\ & - \tfrac{1}{6} C_{\alpha \beta,\gamma \delta} F_{\alpha \beta} F_{\gamma \delta} - \ldots
\end{split}
\end{equation}
Here, $\mu_\alpha$ and $\Theta_{\alpha \beta}$ are the permanent dipole and quadrupole moments, respectively. Coefficients $\alpha$, $\beta$ and $\gamma$ are static dipole polarizabilities of different order, and they have been treated earlier \cite{Tiihonen2016a}. $A$, $B$ and $C$ are called dipole--quadrupole, dipole--dipole--quadrupole, and quadrupole--quadrupole polarizabilities, respectively, and they are the main focus of this article.

The derivation of field-free estimators is done in the spirit of the Hellman--Feynman theorem: we can solve for the polarizabilities by differentiating with respect to the perturbation in the zero-field limit. The differentiation of a diagonal observable is straightforward, and it is explained in more detail in our previous work \cite{Tiihonen2016a}. Nevertheless, for field-gradient polarizabilities this results in
\begin{widetext}
\begin{align}
A_{\alpha\gamma,\alpha\beta}
&= -3\lim_{F \rightarrow 0} \frac{\partial}{\partial F_{\alpha\beta}} \frac{\partial}{\partial F_\gamma} E^{(1)}  
= 3\lim_{F \rightarrow 0} \frac{\partial}{\partial F_{\alpha\beta}} \mu_\gamma = \beta \left[ \langle \tilde \Theta_{\alpha\beta} \tilde \mu_\gamma \rangle - \langle \tilde \Theta_{\alpha\beta} \rangle \langle \tilde \mu_\gamma \rangle \right],
\label{eq-A}
\\ 
B_{\alpha\beta,\gamma\delta} &= 
 -3 \lim_{F \rightarrow 0} \frac{\partial}{\partial F_{\alpha_\beta}} \frac{\partial}{\partial F_{\gamma}} \frac{\partial}{\partial F_{\delta}} E^{(1)} 
= 3 \lim_{F \rightarrow 0} \frac{\partial}{\partial F_{\gamma\delta}} \alpha_{\alpha \beta}
\label{eq-B}
\\ \nonumber & = \beta^2 \left[ \langle \tilde \Theta_{\alpha\beta} \tilde \mu_\gamma \tilde \mu_\delta \rangle + 2\langle \tilde \Theta_{\alpha \beta} \rangle\langle \tilde \mu_\gamma \rangle\langle \tilde \mu_\delta \rangle -  \langle \tilde \Theta_{\alpha\beta} \tilde \mu_\gamma \rangle\langle \tilde \mu_\delta \rangle - \langle \tilde \Theta_{\alpha\beta} \tilde \mu_\delta \rangle \langle \tilde \mu_\gamma \rangle \right],
\\
C_{\alpha\beta,\gamma\delta} 
&=  -3 \lim_{F \rightarrow 0} \frac{\partial}{\partial F_{\alpha\beta}} \frac{\partial}{\partial F_{\gamma\delta}} E^{(1)} 
= \lim_{F \rightarrow 0} \frac{\partial}{\partial F_{\alpha\beta}} \Theta_{\gamma\delta} 
= \tfrac{\beta}{3} \left[ \langle \tilde \Theta_{\alpha \beta} \tilde \Theta_{\gamma \delta} \rangle - \langle \tilde \Theta_{\alpha\beta}\rangle \langle \tilde \Theta_{\gamma\delta} \rangle \right],
\label{eq-C}
\end{align}
\end{widetext}
where $\beta=1/k_BT$ is the inverse temperature. We stress that the correct order of computation in the PIMC algorithm is the following: the properties marked with $\sim$, \emph{i.e.} $\tilde \mu$ and $\tilde \Theta$, must first be averaged over a single trajectory before any multiplication inside angle-brackets. Estimates for the field-gradient induced polarizabilities can be made once the actual observables, such as $\langle \tilde \Theta_{\alpha\beta} \tilde \mu_\gamma \rangle$, are obtained in a reasonable precision.


\section{Results}
We demostrate the finite temperature computation of the field-gradient polarizabilities with our path-integral Monte Carlo code. Besides the new estimators from Eqs. \eqref{eq-A}--\eqref{eq-C}, the technical details of the method are described elsewhere, \emph{e.g.} in Refs.~\cite{Ceperley1995,Kylanpaa2011a,Tiihonen2016a}. With only up to two electrons (or positrons), we can assume opposite spins and avoid the Fermion sign problem. This gives exact \emph{boltzmannon} statistics and a very small error from finite imaginary time-step $\tau$. However, just to be sure we carry out the simulations with several different time-steps and then extrapolate to $\tau\rightarrow 0$. For adiabatic simulations including particles with $Z>1$, \emph{i.e.} Helium, Lithium or Beryllium nuclei, we use time-steps $\tau=0.0125, 0.025, 0.05$; otherwise $\tau=0.025, 0.05, 0.1$. Total energies are extrapolated quadratically, but polarizabilities linearly. The statistical error estimate is given by standard error of the mean (SEM) with 2$\sigma$, \emph{i.e.}~2SEM. All results are given in atomic units.

\begin{table}[!hb]
\begin{threeparttable}
\caption{Total energies $E$, dipole--dipole--quadrupole polarizabilities $B$ and quadrupole--quadrupole polarizabilities $C$ of atomic systems with fixed nucleus, matched with suitable literature references. \label{tbl-adi1} }
\begin{tabular}{ p{0.8cm}p{2.4cm}p{2.4cm}p{2.4cm}} \hline \hline
		& $E$				& $B_{zz,zz}$ 			& $C_{zz,zz}$ \\ \hline
& \\
H		& $-0.49995(3)^a$	& $-106.5(3)^a$			& 5.003(4)$^a$	 \\
		& $-0.5$			& $-106.5^b$			& 5.0$^b$ \\
& \\
He		& $-2.9032(2)^a$	& $-7.37(9)^a$			& 0.819(3)$^a$ \\
		& $-2.90372^c$		& $-7.3267^d$			& 0.8150$^d$		\\
& \\
Li$^+$	& $-7.2797(9)^a$	& $-0.122(4)^a$			& 0.0381(13)$^a$ \\
		& $-7.279913^e$		& $-0.1214^d$			& 0.03796$^d$  \\
& \\
Be$^{2+}$&$-13.6478(9)^a$	& $-0.00853(19)^a$		& 0.005144(10)$^a$ \\
		& $-13.655566^e$	& $-0.008393^d$			& 0.0051067$^d$  \\
\hline \hline
\end{tabular}
\begin{tablenotes}
\item $^a$This work, $^b$Bishop \emph{et al.} \cite{Bishop1995}, $^c$Nakashima \emph{et al.} \cite{Nakashima2007}, $^d$Bishop \emph{et al.} \cite{Bishop1989}, $^e$Johnson \emph{et al.} \cite{Johnson1996}
\end{tablenotes}
\end{threeparttable}
\end{table}

\begin{table*}[t]
\begin{threeparttable}
\caption{Total energies $E$, independent quadrupole moments $\Theta$, dipole--dipole--quadrupole polarizabilities $B$ and quadrupole--quadrupole polarizabilities $C$ of molecular systems at fixed orientation, matched with suitable literature references. \label{tbl-adi2} }
\begin{tabular}{ p{0.9cm}p{2.1cm}p{1.9cm}p{1.7cm}p{1.7cm}p{1.6cm}p{1.6cm}p{1.6cm}p{1.6cm}p{1.6cm} } \hline \hline
		& $E$				& $\Theta_{zz}$		& $B_{zz,zz}$	& $B_{xx,xx}$	& $B_{xx,zz}$	& $B_{xz,xz}$	
& $C_{zz,zz}$	& $C_{xx,xx}$	& $C_{xz,xz}$ \\ \hline

H$_2^+$	& $-0.6026(2)^a$	& 1.53063(8)$^a$	& $-41.9(8)^a$	& $-13.25(13)^a$& 7.31(22)$^a$	& $-18.10(4)^a$
& 1.911(12)$^a$	& 1.267(5)$^a$	& 1.1945(7)$^a$ \\
		& $-0.602634^b$		& 1.5307$^c$		& $-41.869^d$	& $-13.249^d$	& 7.3052$^d$	& $-18.099^d$	
& 1.9113$^d$	& 1.2670$^d$	& 1.1945$^d$ \\

H$_2$	&$-1.17419(27)^a$	& 0.4563(2)$^a$		& $-91.1(6)^a$	& $-66.9(6)^a$	& 34.4(6)$^a$	& $-59.0(3)^a$
& 6.00(2)$^a$	& 4.93(1)$^a$	& 4.185(6)$^a$ \\
		&$-1.174474^e$		& 0.45684$^f$		& $-90.29^g$	& $-66.83^g$	& 34.37$^g$		& $-59.00^g$	
& 5.983$^g$		& 4.927$^g$		& 4.180$^g$ \\

H$_3^+$	& $-1.3438(3)^a$	& $-0.91947(8)^a$	& $-11.7(2)^a$	& $-19.1(2)^a$	& 9.1(3)$^a$	& $-11.08(3)^a$	
& 1.557(8)$^a$	& 2.079(6)$^a$	& 1.2446(8)$^a$ \\
		& $-1.3438356^h$	& $-0.9293^i$		& 				& 				& 				& 	
& 				& 				&  \\

HeH$^+$	& $-2.976(1)^a$		& 1.24950(17)$^a$	& $-5(10)^a$	& $-2.05(14)^a$	& 1.0(3)$^a$	& $-2.24(11)^a$
& 0.59(2)$^a$	& 0.397(8)$^a$	& 0.3384(7)$^a$ \\
		& $-2.978706^j$		&					& 				& 				& 				& 	
& 				& 				&  \\
\hline \hline
\end{tabular}
\begin{tablenotes}
\item $^a$This work, $^b$Turbiner \emph{et al.} \cite{Turbiner2011}, $^c$Bates \emph{et al.} \cite{Bates1953}, $^d$Bishop \emph{et al.} \cite{Bishop1979}, $^e$Kolos \emph{et al.} \cite{Kolos1968}, $^f$Poll \emph{et al.} \cite{Poll1978}, $^g$Bishop \emph{et al.} \cite{Bishop1991}, $^h$Turbiner \emph{et al.} \cite{Turbiner2013}, $^i$Borkman \cite{Borkman1971}, $^j$Pachucki \cite{Pachucki2012}
\end{tablenotes}
\end{threeparttable}
\end{table*}

In the following, we present polarizability data and discussion for a variety of isolated one or two-electron systems: H, H$^-$, Li$^+$, Be$^{2+}$, H$_2^+$, H$_2$, Ps$_2$, H$_3^+$ and HeH$^+$. We run two kinds of simulations: adiabatic and non-adiabatic. In the adiabatic, or Born--Oppenheimer approximation (BO), the nuclei are fixed in space, reducing symmetry and producing various directional components to polarizabilities. The adiabatic approximation inhibits the rovibrational motion, and thus, at reasonably low temperatures the difference to absolute zero is negligible. Therefore, we start by establishing the validity of our method by comparing our BO results to the available 0 K reference data.

Excellent summary of independent tensorial polarizabilities for each point group is given in Ref.~\cite{Buckingham2007}. In Table~\ref{tbl-adi1}, we present BO results for all of the spherically symmetric systems: $B_{zz,zz}$, $C_{zz,zz}$ and the total energy $E$. Furthermore, the results for the molecular systems, i.e. H$_2^+$, H$_2$ H$_3^+$ and HeH$^+$, are given in Table~\ref{tbl-adi2}. Each molecular system has one independent quadrupole moment $\Theta_{zz}$ and four independent dipole--dipole--quadrupole polarizabilities: $B_{zz,zz}$, $B_{xx,xx}$, $B_{zz,xx}$, and $B_{xz,xz}$. Similarly, there are three independent components of quadrupole--quadrupole polarizabilities: $C_{zz,zz}$, $C_{xx,xx}$, $C_{xz,xz}$. Distinct symmetries also lead to a few non-zero dipole--quadrupole polarizabilities $A$: for H$_3^+$, $A_{y,yy}= -0.653(7)$ and for HeH$^+$ $A_{z,zz}=-0.48(6)$ and $A_{x,zx}=-0.0657(10)$. The principal axis $z$ is by default the line connecting the two nuclei, but for triangular H$_3^+$ it is perpendicular to the plane of protons. In BO simulation the molecules are placed at the equilibrium geometries, namely $R_{\mathrm{H_2^+}}=2.0$, $R_{\mathrm{H_2}}=1.4$, $R_{\mathrm{H_3^+}}=1.65$ and $R_{\mathrm{HeH^+}}=1.46$. The dipole and quadrupole moments are calculated with respect to the center-of-mass. The temperature was set to $T=2000$ K, which still corresponds to the electronic ground state for most neutral and positively charged systems. Still, in the last decimals of the total energy, a small thermal increment can be observed. Besides that, the agreement is good with all of the available 0 K literature references \cite{Bishop1995,Nakashima2007,Bishop1989,Turbiner2011,Bates1953,Bishop1979,Kolos1968,
Poll1978,Bishop1991,Turbiner2013,Borkman1970,Pachucki2012}.

\begin{table}[b]
\begin{threeparttable}
\caption{Total energies $E$, dipole--dipole--quadrupole polarizabilities $B$ and quadrupole--quadrupole polarizabilities $C$ of H$^-$, PsH and Ps$_2$ with the protons fixed but the positrons free. The values have been extrapolated to $T\rightarrow 0$ and matched with literature references, where available. \label{tbl-ps} }
\begin{tabular}{ p{0.8cm}p{2.4cm}p{2.4cm}p{2.4cm}} \hline \hline
		& $E$				& $B_{zz,zz}$ 				& $C_{zz,zz}$ \\ \hline
H$^-$	& $-0.52777(11)^a$	& $-4.8(5)\times 10^{5a}$	& 2572(85)$^a$\\
		& $-0.52775^b$		& $-4.843\times 10^{5c}$	& 2591.6$^c$  \\

PsH		& $-0.7893(2)^a$	& 5270(190)$^a$				& 260(3)$^a$ \\
		& $-0.78913^d$		& 							& 	  \\

Ps$_2$	& $-0.51593(6)^a$	& 0(330)$^f$				& 447(10)$^a$ \\
		& $-0.5160038^e$	& 							& \\ \hline \hline
\end{tabular}
\begin{tablenotes}
\item $^a$This work, Lin \cite{Lin1995}; $^b$Nakashima \emph{et al.} \cite{Nakashima2007}, $^c$Pipin \emph{et al.} \cite{Pipin1992}, $^d$Frolov \emph{et al.} \cite{Frolov1997}, $^e$Bubin \emph{et al.} \cite{Bubin2007}
\item $^f$This work; estimating anything other than 0 is unfeasible because of the large fluctuations.
\end{tablenotes}
\end{threeparttable}
\end{table}

\begin{table}[b]
\begin{threeparttable}
\caption{Total energies, dipole--dipole--quadrupole polarizabilities and quadrupole--quadrupole polarizabilities extrapolated to 0 K. Quadratic fit is used for $E$, and Eq.~\eqref{eq-fit} with optimal $x$ for $B$ and $C$. \label{tbl-nonadi} }
\begin{tabular}{ p{0.8cm}p{2.4cm}p{2.4cm}p{2.4cm}} \hline \hline
		& $E$				& $B_{ZZ,ZZ}$ 				& $C_{ZZ,ZZ}$ \\ \hline
H$_2^+$	& $-0.596(2)^a$		& 3000(850)$^a$				& 580(150)$^a$ \\
		& $-0.597139^b$		& 							&  \\
& \\
H$_2$	& $-1.1625(11)^a$	& 160(35)$^a$				& 32(6)$^a$ \\
		& $-1.164025^c$		& 							&  \\
& \\
H$_3^+$	& $-1.323(5)^a$		& 860(720)$^a$				& 157(39)$^a$ \\
		& $-1.313568^d$		& 							&  \\
& \\
HeH$^+$	& $-2.9670(8)^a$	& 3.4(1.7)$\times10^{6a}$	& 406(110)$^a$ \\
		& $-2.96627^e$		& 							& \\
\hline \hline
\end{tabular}
\begin{tablenotes}
\item $^a$This work (extrapolated to 0 K), $^b$Tang \emph{et al.} \cite{Tang2014}, $^c$Stanke \emph{et al.} \cite{Stanke2008}, $^d$Kyl\"anp\"a\"a \emph{et al.} \cite{Kylanpaa2010}, $^e$Calculated based on Refs. \cite{Pachucki2012} and \cite{Tung2012}
\end{tablenotes}
\end{threeparttable}
\end{table}

\begin{figure*}[ht!]
\includegraphics[width=\textwidth]{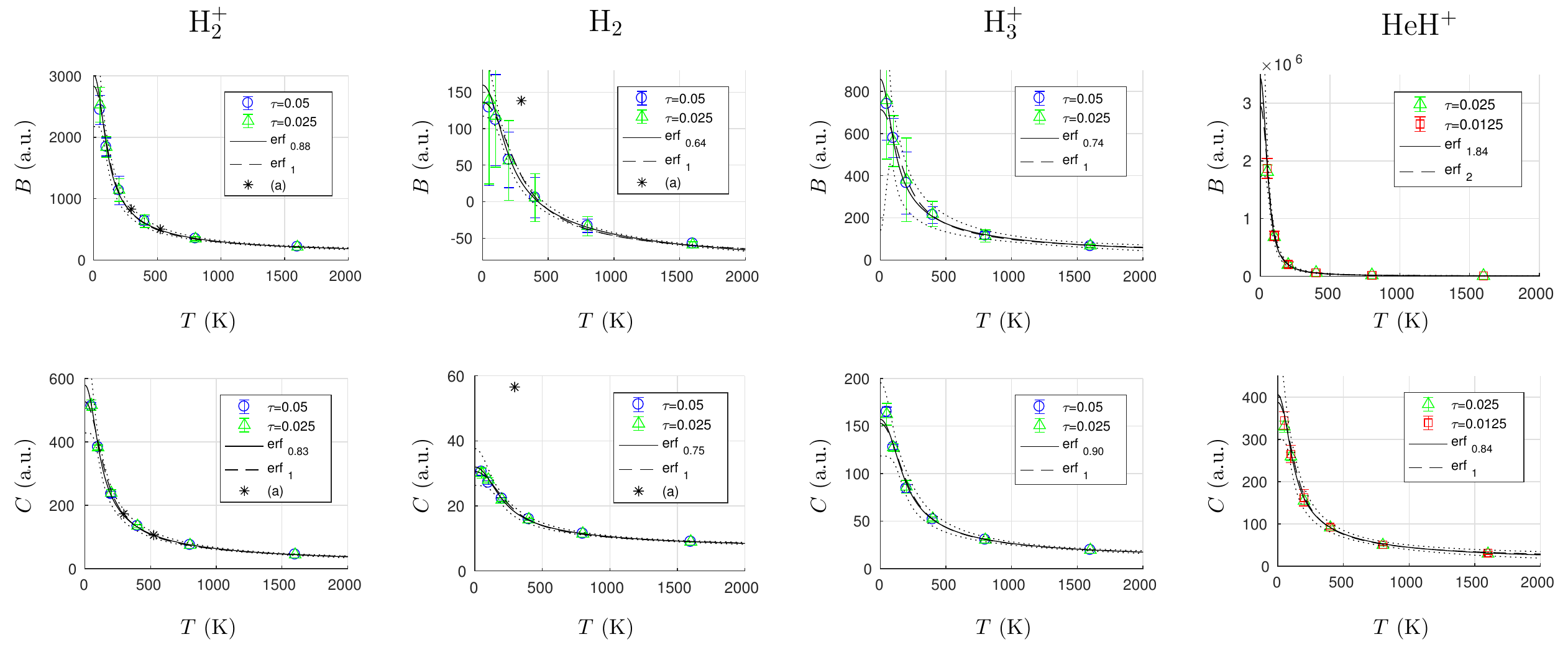}
\caption{Rovibrationally averaged dipole--dipole--quadrupole polarizabilities $B_{ZZ,ZZ}$ and quadrupole--quadrupole polarizabilities $C_{ZZ,ZZ}$ for nonadiabatic simulations of molecular systems plotted in different temperatures. A few data points from Ref.~ \cite{Bishop1988} have been marked with (a). Fits to Eq.~\eqref{eq-fit} are presented with solid line for optimal exponent $x$ and dashed line for integer exponent. Dotted lines are 95\% confidence boundaries given by the fitting algorithm. \label{fig-nonadi} }
\end{figure*}

Sampling the ground state is not as simple for loosely bound H$^-$ and for positronic systems PsH and Ps$_2$. Essentially, the systems need to be simulated at several lower temperatures, \emph{e.g.}, below $T=500$ K, and the data be extrapolated to $T\rightarrow 0$. Values for $E$,  $B_{zz,zz}$ and $C_{zz,zz}$ are presented in Table~\ref{tbl-ps}. All systems share spherical symmetry, but the simulation of Ps$_2$ is not adiabatic, \emph{per se}, since the positrons are fully delocalized. The data for positronium, Ps, is missing because the symmetry of masses $m_{\bar{e}}=m_e$ makes its quadrupole moment vanish. Overall, match is good with the available literature references \cite{Lin1995,Pipin1992,Frolov1997,Bubin2007}. The field-gradient polarizabilities for positron systems have not been published before.

To non-adiabatic simulations we refer as \emph{all-quantum} (AQ), since they include all rovibrational and electronic quantum effects. Thus, we only use it to study the systems whose polarizabilities show considerable thermal coupling, \emph{i.e.}, molecules. Where relevant, we use $m_p=1836.15267248 m_e$ for proton mass and $m_{\mathrm{He}}=7294.2995363 m_e$ for that of He-nucleus. The AQ simulations are done in the laboratory coordinates, which is denoted by capital $Z$. The results are exact rovibrationally averaged quantities and therefore spherically symmetric. Consequently, $A_{ZZ,Z}$ are zero for all systems. The resulting temperature-dependent data for $B_{ZZ,ZZ}$ and $C_{ZZ,ZZ}$ for H$_2^+$, H$_2$, H$_3^+$ and HeH$^+$ are presented in Fig.~\ref{fig-nonadi} in order to show that any time-step effects are negligible. The actual numerical and extrapolated data can be found in the Supplementary material.

Any non-zero electric moments of a quantum system couple to its rotational states, and then this coupling is manifested in the rotational parts of higher order polarizabilities. In high temperatures, this rotational coupling is proportional to the inverse temperature, which has already been proposed \cite{Bishop1988,Bishop1990} and demonstrated \cite{Tiihonen2016a}. Now, for homonuclear molecules H$_2^+$, H$_2$, H$_3^+$ the first non-zero electric moment is the quadrupole moment $\Theta$, and thus, all of these systems show $\sim 1/T$ decay on $B$ and $C$. For HeH$^+$ with non-zero dipole moment $\mu$, the dipole polarizability $\alpha$ is also affected by the coupling \cite{Tiihonen2016a}. Thus, it makes sense that $B$ of HeH$^+$, involving both $\alpha$ and $\Theta$, is in fact proportional to $\sim 1/T^2$.

\begin{figure}[b!]
\includegraphics[width=8.0cm]{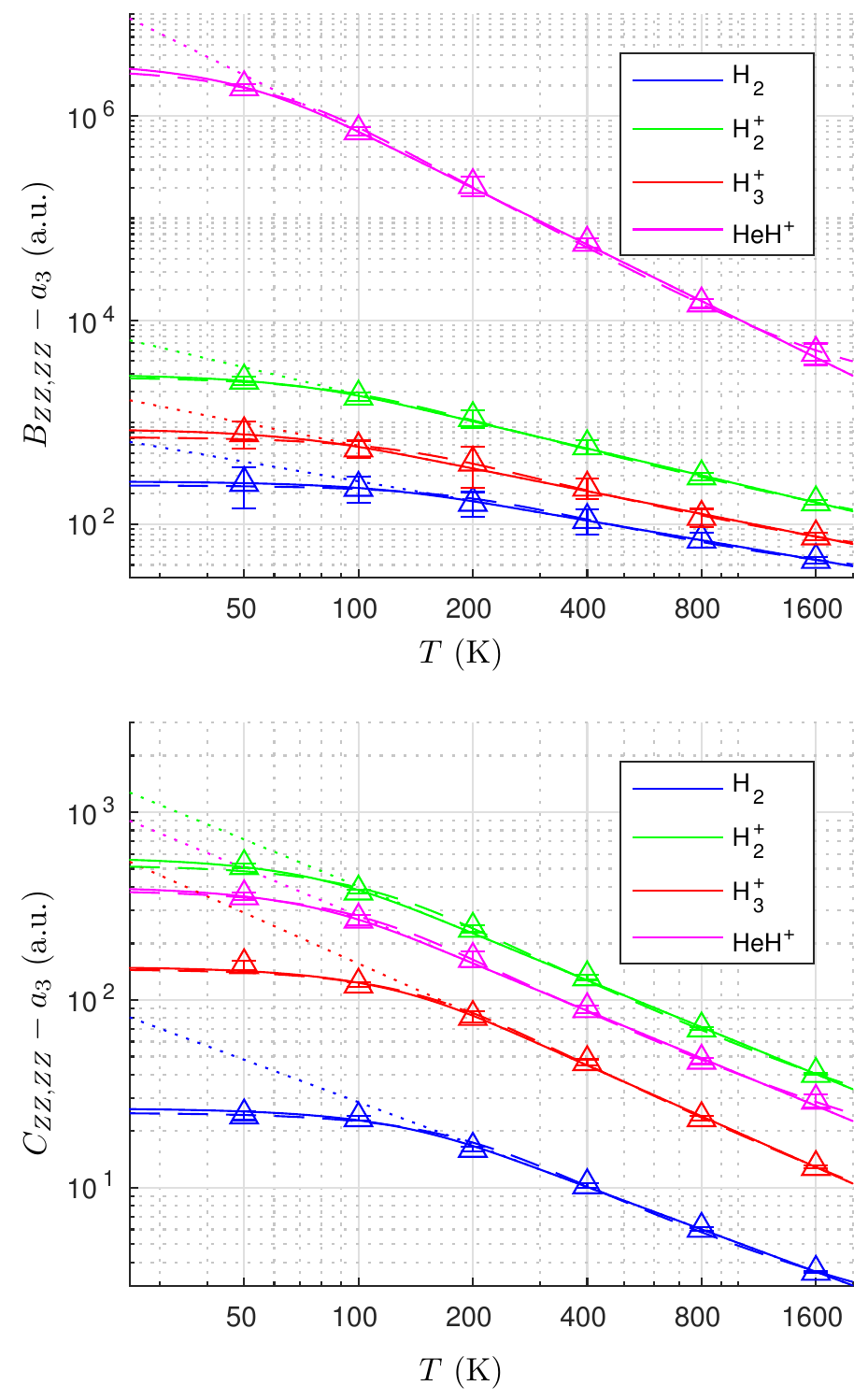}
\caption{Time-step extrapolated data and nonlinear fits for $B$ and $C$ of H$_2^+$, H$_2$, H$_3^+$ and HeH$^+$ on logarithmic scale. The fits to Eq.~\eqref{eq-fit} are done with $x=1$ (dashed) or the optimal $x$ (solid). Dotted lines show the effect of replacing the error function with unity. \label{fig-log} }
\end{figure}

However, the rotational polarizabilities do not diverge in low temperatures, because it takes some energy to activate the rotational states. To model the temperature dependence of the total $B$ and $C$, we propose an \emph{ad hoc} nonlinear function of the form
\begin{equation}
f(T) = \left(\frac{a_1 \cdot \mathrm{erf}(a_2 T)}{T} \right)^x + a_3,
\label{eq-fit}
\end{equation}
where $a_1$, $a_2$ and $a_3$ are coefficients, and the error function $\mathrm{erf}(y)$ is used to saturate the values in a robust way as $T\rightarrow 0$. As argued earlier, a natural choice for the characteristic exponent describing the rotational coupling is $x=1$ ($x=2$ for $B$ of HeH$^+$). However, we also present $x$ optimized by the root-mean-squared error (RMSE) as a crude means of considering nontrivial thermal effects originating from the electronic and vibrational polarizabilities. Nonlinear fitting to time-step extrapolated data has been done using \verb#fitnlm# function in Matlab, which also provides 95\% confidence intervals. Inversed squares of SEM estimates of the PIMC data were used as a weights.

Extrapolation of Eq.~\eqref{eq-fit} to $T=0$ is given by $\tfrac{2}{\sqrt{\pi}} a_1 a_2 + a_3$. The corresponding data for $B$ and $C$ is presented in Table~\ref{tbl-nonadi} together with quadratically extrapolated total energies and appropriate references \cite{Tang2014,Stanke2008,Kylanpaa2010,Pachucki2012,Tung2012}. The raw data and the fitting coefficients can be found in the Supplementary material. Besides Fig.~\ref{fig-nonadi}, the fitted curves are presented on logarithmic scale in Fig.~\ref{fig-log}. It is easier to see that the rotational polarizability is saturated at low $T$ but decays as $~T^{-x}$ as the rotational states get activated. Also, it can be observed that the magnitudes of the rotational parts of $B$ (except for HeH$^+$) and $C$ are clearly in the same order as the corresponding lower order moments, $\Theta_{zz}$, from Table~\ref{tbl-adi2}.

The high-temperature limit of the fit is given by $a_3$. It gives the ballpark of the sum of the vibrational and electronic polarizabilities. Their thermal coupling is much smaller but not negligible. This is manifested in the characteristic exponent $x$: the optimal $x$ in a least-squares fit appears to be slightly smaller than a natural integer, 1 or 2. While the exponent in $T^{-x}$ is probably not the most natural way to model this, it shows evidence on how the vibrational and electronic parts compensate on the decay of rotational polarizability. However, we omit trying to further analyze these nontrivial effects within our scheme.

As a final remark, we discuss the only explicit reference for the finite temperature total polarizabilities given by Bishop \emph{et al.} \cite{Bishop1988}. As shown in Fig.~\ref{tbl-adi1}, their results are a good match for H$_2^+$ but severely overestimated for H$_2$. We suggest that this is caused by inaccuracy of the vibrational wave function basis used by the authors. Due to the electronic correlations, their ground state is not exact, but rather an uncontrollable mixture involving higher excited vibrational eigenstates. According to their own tables, such vibrational bias leads to unintended overestimation of properties, which can be substantial in case of polarizabilities. This example discloses the inherent sensitivity of estimating higher order electric properties in many-body systems.

\section{Summary.}
As a natural continuation to our previous work, we present a scheme to estimate static field-gradient polarizabilities in a field-free PIMC simulation. We apply it on a range of small atoms, ions and molecules, namely H, H$^-$, He, Li$^+$, Be$^{2+}$, Ps$_2$, PsH, H$_2^+$, H$_2$, H$_3^+$ and HeH$^+$. The simulations with the adiabatic approximation and equilibrium geometries are done in the low temperature limit, and they indeed agree well with the 0 K literature references. However, we do not try to push the limits of statistical precision in this study, but rather, we want to give an ample demonstration of our method.

With the given set of systems, the variation in dielectric properties is already large. For instance, H$^-$ or PsH are very diffuse compared to the heavier ions, Li$^+$ and Be$^{2+}$. On the other hand, HeH$^+$ has a permanent dipole moment, and thus, much more diverse dielectric response than the homonuclear molecules. We want to emphasize that all these properties were obtained with the same PIMC procedure varying nothing else than the fundamental properties of the particles.

One of the most advantageous treats of the PIMC method is the exact simulation of the canonical ensemble. Molecules have geometrical anisotropy, and thus, permanent dipole or quadrupole moments, which then reflect in the higher order rotational polarizabilities. Our data indicates that the rotational parts of $B_{ZZ,ZZ}$ and $C_{ZZ,ZZ}$ are dominant at low temperatures, but decay drastically when the temperature is increased. The latter effect has been anticipated in the literature \cite{Bishop1990}, but even our overly simplistic model in Eq.~\eqref{eq-fit} shows that there is plenty of room for improvement. Indeed, the requirements of explicit correlations and non-adiabatic thermal averaging render results of this kind very scarce. By this work, we are hoping to inspire change to that.

\section{Acknowledgements.}
We thank Jenny and Antti Wihuri Foundation and Tampere Univesity of Technology for financial support. Also, we acknowledge CSC--IT Center for Science Ltd. and Tampere Center for Scientific Computing for providing us with computational resources.

\end{document}